# Will AI be overconfident about academic research findings when reliant on abstracts?

Mike Thelwall, University of Sheffield, UK

Large Language Models (LLMs) like ChatGPT, DeepSeek and Gemini seem to be increasingly used for knowledge discovery, information retrieval, and knowledge summaries, including for academic topics. This can result in users being misled, such as due to hallucinations. These problems may be exacerbated for academic knowledge if LLMs base their answers on journal article abstracts when they lack full text access. To test whether the information content of abstracts can be misleading, full text articles were submitted to the GPT-OSS 120B, an LLM from OpenAI, asking it to assess separately the strength the claims for the main result in the abstract, discussion, and conclusion. Outside the social sciences and humanities, claims tended to be stronger in the abstract and conclusions than the discussion, suggesting that relying on the strength of claims in abstracts would be misleading. Thus, if LLMs ingest abstracts but not full texts, there is a risk that they will be overconfident about the findings and pass it on to users in response to relevant prompts. This is another reason to be cautious about using LLMs for academic-related knowledge discovery and summaries.

## Introduction

Historically, students, researchers and end users learned academic knowledge from textbooks or reading journal articles in libraries. This progressed to digital libraries and desktop access, including through keyword searching (Hemminger et al., 2007). With the advent of Google and Google Scholar, finding scholarly information became partly algorithmically determined (Herrera, 2011; López-Cózar et al., 2017), but ultimately the learner needed to read the original document, or perhaps a human-authored review article or Wikipedia article summary, to discover the knowledge. If they chose to just read the abstract, then this would be unprofessional but at least they would know that they were working from incomplete information.

Since the Google Scholar era, Large Language Models (LLMs) have become part of the academic landscape, with a 2024 survey finding that 96% expected AI to speed knowledge discovery (Elsevier, 2024). They are also part of the literature search and reviewing process for many scholars and are also commonly used (e.g., already 36% in 2023: Van Noorden & Perkel, 2023) as an alternative to reading the source or a human-authored summary. This has occurred despite the known risk of hallucinations (e.g., Cheng et al., 2026). A researcher might now query, “Which statistical test is appropriate for comparing Likert scale responses from three different groups” and receive an LLM explanation, perhaps with references. This is efficient since it avoids the need to identify the appropriate search terminology (e.g., non-parametric statistics, although this is very broad) and then a relevant knowledge source (e.g., a statistics textbook or journal article covering multigroup non-parametric ordinal comparisons). In addition, the LLM explanation is likely to be tailored to the use case asked about, giving easy to understand knowledge and advice. LLMs also seem to be very persuasive, so the user may not feel the need to invest the considerable time needed to verify the information from original sources. If this leads to at least a partial reliance on LLMs as an alternative to human-authored sources, then it is important to investigate contexts in which the LLM responses can be misleading. This article focuses on whether LLMs understand the certainty with which academic findings are reported.

Given the use of LLMs for information discovery, it is unfortunate that copyright laws and paywalls do not always permit an LLM to ingest knowledge from the source unless they pay a licence fee (Kwon, 2024). Thus, a LLM might know about knowledge in a journal article from its abstract or a summary blog post rather than its full text. When research is mediated in this way, the carefulness with which the original authors have caveated their results may be lost. Although it is known from previous studies that academics use careful linguistic strategies to express an appropriate degree of confidence in their results (Hyland, 1998, 2005), no previous study has investigated the relationship between the source of academic information and the certainty with which it is expressed. This article partly fills this gap by investigating whether academic results are described with differing degrees of certainty in three key sections of a paper: the abstracts, conclusion, and discussion. The purpose is to assess whether LLMs would learn a misleading degree of certainty if they only ingested abstracts. This could be thought of as a weak hallucination: not stating a false fact but misrepresenting the strength of evidence for it. As background to this, field differences are investigated. The following research questions drive the study.

1. RQ1: Are there differences between fields in the certainty with which results are expressed?
2. RQ2: Are results expressed with different degrees of certainty in abstracts, discussions, and conclusions?

## Background: Expressing degrees of uncertainty in academic writing

Except for pure mathematics and symbolic logic, academic research findings are rarely established without doubt (Strevens, 2020). Opinions about whether a fact is true can change over time, including if new theories fit the data better, or new evidence undermines a previous interpretation. Moreover, researchers typically must make simplifying assumptions and may focus on phenomena that they have relatively convenient access to, limiting generalisation. For individual claims, three aspects can be broadly distinguished (Hyland, 2005).

- Epistemic certainty: the extent to which a claim is supported by evidence. This may change after publication as alternative claims/theories emerge.
- Linguistic certainty: the strength with which a claim is made in an academic publication.
- Social certainty: the extent to which a claim is accepted by other researchers. This may change along with epistemic certainty changes or due to social factors.

For individual claims in new academic research, scholars must understand the epistemic certainty of their claims when deciding on the level of linguistic certainty to express, presumably with the end goal of increasing the claim's social certainty, whether in terms of persuading editors/reviewers or subsequent readers. Overclaiming, in the sense of expressing too much confidence in results, may lead to articles being rejected or readers questioning the researcher's skill. Making appropriate claims therefore requires a combination of subject expertise, linguistic fluency, and social/cultural knowledge (Bazerman, 1988; Hyland, 2005).

### *Linguistic strategies to express varying degrees of certainty*

Academic authors can express their desired level of confidence in their claims partly by using linguistic hedges like *might* to indicate uncertainty or linguistic boosters like *proven* for certainty (Hyland, 1998, 2005). Although not unique to academic writing (e.g., Kuik, 2023), hedging seems to be the standard approach. As reviewed below, there have been many studies of hedging in academic writing, typically either to understand it or to help non-native English speakers to write appropriately for publication. Despite this, all previous studies have been restricted to one or a few journals. This is unfortunate because knowledge of how the phenomenon varies between fields is essential if authors are to be guided appropriately.

There are some international differences in the use of hedges in academic writing. A comparison of hedges and boosters in English language teaching journal articles between Turkish and native English speakers, found few differences overall although Turkish speakers used more modal auxiliaries and semi-modal verbs (e.g., could) whereas English speakers used other verbs more (e.g., believe) as well as a wider range of hedges (Tables 6 to 12 of: Demir, 2018). Native English speakers also used a wider range of hedges than Serbian authors of tourism and hospitality research (Radovanović & Vojnović, 2023). For foreign studies journal articles, hedges were more commonly used by native English speakers than by native Vietnamese speakers (Nguyen Thi Thuy, 2018). Similar findings have occurred in other contexts (Rezanejad, et al., 2015). Hedges were found to be more common in English than Malay academic texts about educational research (Loi et al., 2019). A two-language comparison of applied linguistics articles found that more hedging was used in German language writing than in Chinese language writing (Yu & Wen, 2022).

There are field differences in the use of hedges. In Slovenian doctoral theses, modal adverbs are more commonly used to express uncertainty about analyses or findings in the social sciences and humanities than in the natural sciences. In contrast, they were more used to describe the subject analysed in the natural sciences (Lenardič, & Fišer, 2021). Even within fields, the use of hedges can vary between topics, as in the *Journal of Applied Linguistics* (Livytska, 2019). A comparison of 150 basic science and clinical science in an Iranian medical journal found that hedges were more used in clinical papers, although the difference was not statistically significant (Hamidi et al., 2025).

There may also be gender differences in hedging. Based on an analysis of 60 articles in chemistry and applied linguistics, female authors used hedges more than males, although the difference was not statistically significant (Mirzapour, 2016). The prevalence of hedges within articles also varies between sections. The largest scale study so far (Yao et al., 2023), used a Python program to count hedging words for 2652 articles published between 1997 and 2021 in the journal Science. It found that the use of hedges had decreased over time, arguing that this increase aligned with an increasingly promotional approach to science writing (e.g., Figure 1 showing the number of hedges per 10,000 words: Yao et al., 2023).

Despite the above research into hedging, only one study has given findings that are directly relevant to the current paper. For Spanish business management journal articles, hedges were most common in the introduction and discussion than in the methods and conclusion sections and were more common in English than Spanish (Mur-Dueñas, 2021).

## Methods

The research design was to obtain a set of large journals from a diverse range of fields, use LLMs to detect their use of boosters and hedges and estimate the certainty of their results, and then compare the results between journals. Whilst the use of LLMs for data processing is usually not recommended without human supervision or checking, for this article's research goal, the important factor is not how humans would interpret the level of certainty expressed in scientific writing, but how LLMs would interpret it. This is because LLMs are the mediator that may subsequently deliver the information to human users. Thus, a comparison between human and LLM perceptions of section strengths is unnecessary.

## *Data*

Journals from MDPI were chosen for the raw data. Apparently uniquely for a science-wide English English-language publisher, MDPI makes a complete set of its articles available in easy to process XML form. XML is easier to extract text from in an error-free way than HTML and PDF so is the optimal format. MDPI has the disadvantage that its journals seem to be topic focused rather than field focused, so are multidisciplinary. This complicates comparisons between fields. Nevertheless, there is not an appropriate alternative science-wide source of full text articles.

The XML of MDPI journals was downloaded on 19 December 2025 and the full text extracted without metadata, references, or appendices. Eight journals were manually selected to represent different areas of academic research, including all major types. Of course, all fields are unique so any selection risks being unrepresentative. For this reason, general journals were chosen for the humanities and social sciences, but this option was not available for other fields. For journals with more than 5000 articles, a random number generator was used to extract a sample of 5000, to avoid overusing LLM queries.

## *Large Language Models*

There are now thousands of LLMs without clear information about which is the best for this type of task, so a combination of the best available evidence and intuition was used to select one. Cloud-based LLMs were not used due to their cost for this task. Instead, the relatively new and large (at the time of writing) GPT-OSS 120B was downloaded and used. Although Gemma-3-27B has good performance on other academic tasks (Thelwall & Mohammadi, 2026), GPT-OSS 120B is substantially newer and larger, originating from another of the main LLM companies, OpenAI. Its much larger size should also help it to cope with the complex task of certainty estimation. Other LLMs could have also been selected for comparison, but the purpose was to identify that differing strengths were possible for LLMs rather than detecting all those with the same issue.

It is important to design effective prompts to elicit clear information from LLMs. For RQ1, a prompt requesting structured information was used to make the extraction of the answers error free. After testing different prompts on small samples, the following was used to extract information about the use of hedges and boosters and the overall claim strength. A ten-point scale was used because previous experience had found this level of granularity worked well.

> You are an academic researcher and concerned with the linguistic style that authors use. You analyse hedges, which are terms like might and should, that imply caution, uncertainty, or modesty in claims. You also analyse boosters, which are terms like clearly and prove, that express certainty, conviction or strong commitment to a claim. Answer the following questions about hedges and boosters in the journal article below, strictly following the given format.
>
> Does the article below use hedges when reporting or discussing its results (y/n)?:

Does the article below use boosters when reporting or discussing its results (y/n)?:
What is the overall strength of the claim for the paper's results on a scale of 1 to 10, where 1 is a very weak claim and 10 is a very strong claim (1-10)?:
####
[full text of article]

For RQ2, a prompt was used to compare the strengths of the main claim between the abstract, discussion, and conclusions, as follows. The prompt start by asking for the main result in the article to reduce the risk that information about the three different sections might refer to different claims.

You are an academic researcher and concerned with how authors linguistically describe their research results in different article sections. Answer the following questions about the strength of the claims for the main results in the journal article below, strictly following the given format.
Does the article below include an abstract, a discussion section and a conclusion section (y/n)?:
What is the main result reported in the article?:
How strongly does the author claim the main result in the article's abstract on a scale of 1 to 10, where 1 is a very weak claim and 10 is a very strong claim (1-10)?:
How strongly does the author claim the main result in the article's discussion section on a scale of 1 to 10, where 1 is a very weak claim and 10 is a very strong claim (1-10)?:
How strongly does the author claim the main result in the article's conclusions section on a scale of 1 to 10, where 1 is a very weak claim and 10 is a very strong claim (1-10)?:
####
[full text of article]

The results were extracted by a bespoke Python program. Articles were ignored when a score was missing. This could occur because the relevant sections were missing. For example, one report included, “How strongly does the author claim the main result in the article's conclusions section on a scale of 1 to 10 (1 = very weak, 10 = very strong)?: N/A (the article does not contain a separate Conclusions section)”.

### *Analysis*

The RQ1 results were reported descriptively and the RQ2 statistically. For RQ2, 95% confidence intervals were calculated using the standard t-test formula since the data is scalar. The data failed Shapiro-Wilk normality tests but the t-test formula is still appropriate since the sample sizes are large and there is limited scope for skewing, given the range 1-10.

## Results

Except for *Healthcare*, most articles in each journal did not have a complete set of results (Table 1). The weakest case was Humanities, presumably because humanities scholars often do not use discussion or conclusion sections. Nevertheless, there were at least 193 qualifying articles in each field.

Table 1. Original and final (with scores for all three sections) sample sizes.

| Journal | Original sample size | Final sample size | Percent valid |
|---|---|---|---|
| *Agronomy* | 5000 | 2432 | 49% |
| *Electronics* | 5000 | 1863 | 37% |
| *Fractal and Fractional* | 3554 | 1091 | 31% |
| *Healthcare* | 5000 | 2955 | 59% |
| *Humanities* | 1217 | 193 | 16% |
| *Molecules* | 5000 | 1472 | 29% |
| *Social Science* | 4302 | 1840 | 43% |
| *Sustainability* | 5000 | 2303 | 46% |

## *RQ1: Hedges and overall strength of claims*

There are substantial field differences in the use boosters and hedges and the average strength of knowledge claims, as judged by GPT-OSS 120B (Figure 1). Whilst hedges to emphasise result uncertainty are close to ubiquitous in most fields, they are less universal in two journals: *Fractal and Fractional*, and *Electronics*. In both cases, this might be due to formal mathematical content being certain and not needing hedges. LLMs could either estimate certainty from linguistic formulations, such as hedges, or may use the topic as an indirect cue. In contrast, booster terms are not near universal in any field, but they are more common when hedges are rarer, suggesting that some authors can use one instead of the other instead of a balance between the two. Finally, claim strengths seem to be weakest in the humanities and social sciences and strongest in formal sciences. This may reflect inherently unpredictable and varied nature of humans (*Humanities*, *Social Science*, *Healthcare*), making certainty intrinsically less attainable.

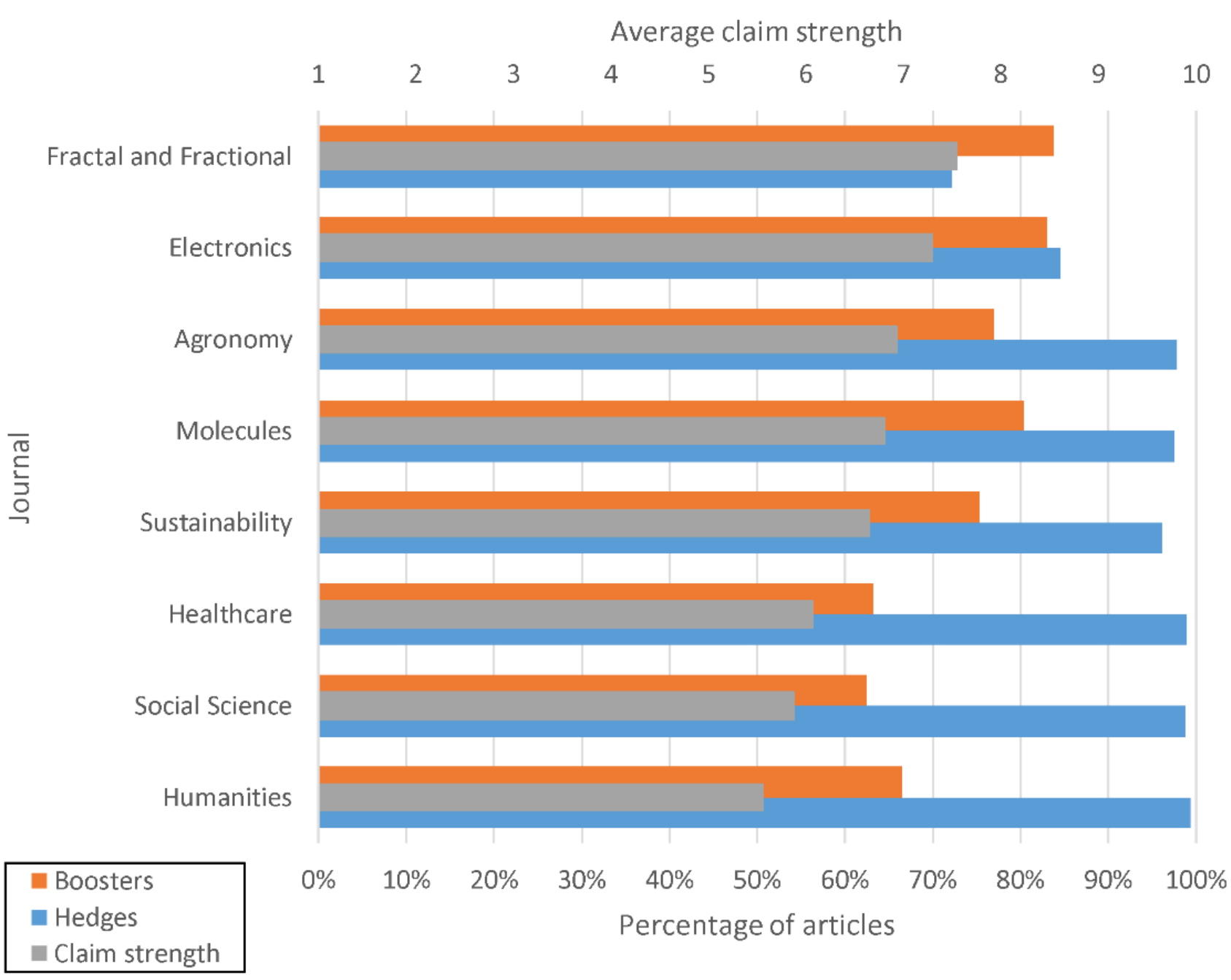

Figure 1. Average overall claim strength and the proportion of articles with hedges and booster terms by journal, as claimed by GPT-OSS 120B. Journals are in order of claim strength.

### *RQ2: Main claim strengths in abstracts, discussion and conclusions*

Although the claimed strength of the main results, according to GPT-OSS 120B, tends to be similar in the abstract, discussion and conclusions for *Humanities* and *Social Sciences*, it differs for all other journals (Figure 2). For the remaining six journals, the confidence of the discussion in the main result is the lowest. The strongest confidence is sometimes in the conclusions and sometimes in the Abstract (e.g., *Healthcare*, *Molecules*). The difference is statistically significant (in the sense of non-overlapping 95% confidence intervals, which indicates p<0.05 for a hypothesis test) in most cases. For *Humanities*, the stronger abstract claims could be an artefact of the small sample size (overlapping confidence intervals) or an underlying different rhetorical strategy.

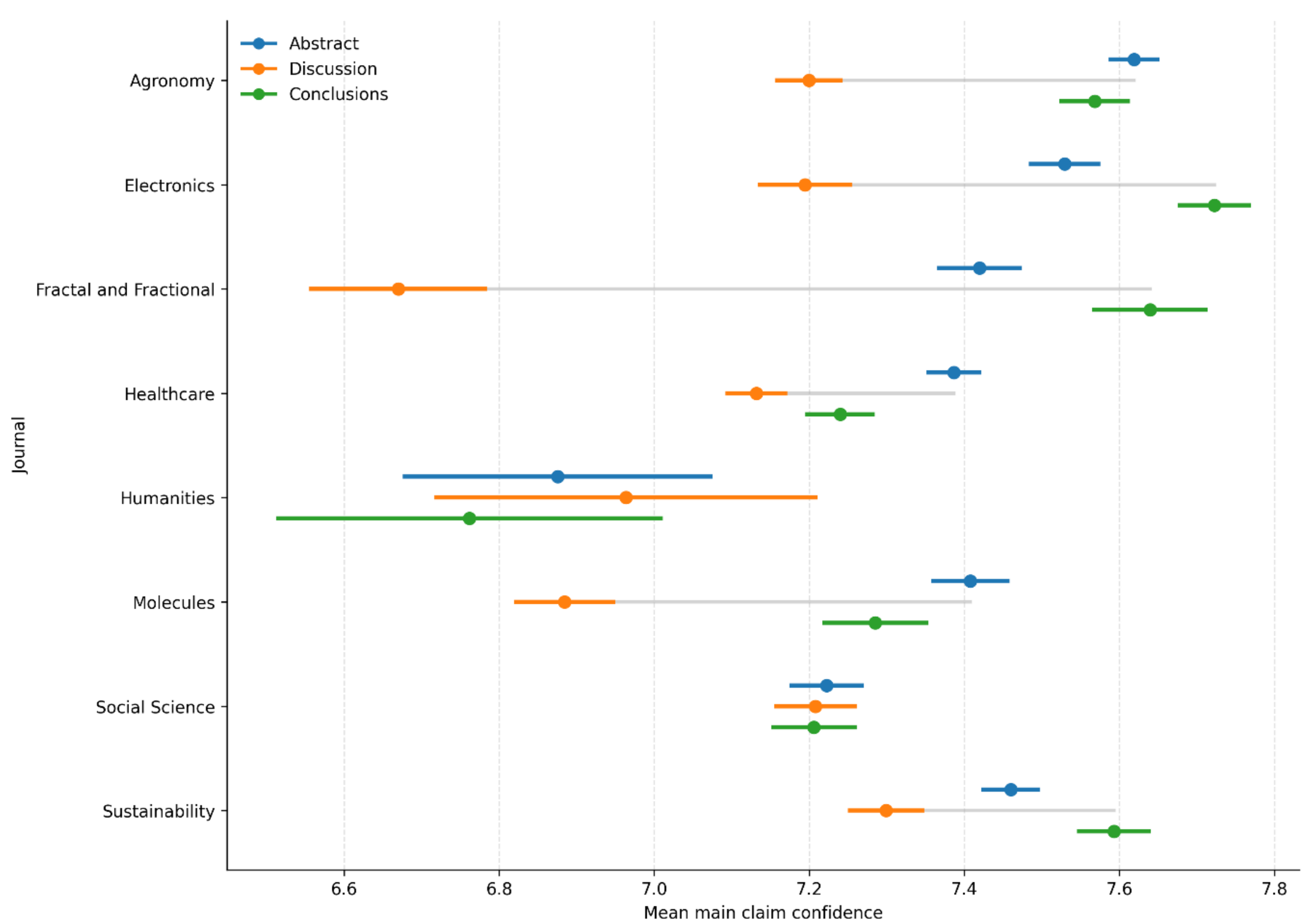

Figure 2. Average main claim strength in the abstract, conclusions and discussion by journal, according to GPT-OSS 120B. Journals are in alphabetical order. Bars indicate 95% confidence intervals. Faint grey bars indicate the spread between sections.

## Discussion and conclusions

This study is limited to a single LLM, a small set of journals, and a single publisher. The results may be different for other LLMs and journals. The finding that the main result is expressed more strongly in the conclusion than the discussion aligns with a previous study of Spanish business journals that found hedging to be more prevalent in the discussion than the conclusions (Mur-Dueñas, 2021).

Overall, however, the results show that in some fields LLMs *can* interpret the strength of the main result of a paper differently – stronger from the abstract than from the discussion (usually) and from the conclusion (sometimes). This suggests that when LLMs are used to summarise information from academic sources and they rely on abstracts because they cannot access full texts, then they may express the results with too much certainty.

In practical terms, this conclusion adds to many previous warnings to carefully check LLM reports (e.g., Cheng et al., 2026), here for the special case of academic knowledge, and especially outside of the social sciences and humanities. Moreover, LLM owners may need to introduce safeguards when processing academic information from abstracts or other summary sources, in case the model expresses too much confidence in its results. In the absence of this, authors may need to be careful that their abstracts convey the appropriate degree of confidence in their findings. Finally, the results give another reason to favour open access licences for academic research, and specifically ones that allow AI processing.

## Acknowledgement

I used ChatGPT 5.2 to write the Python code used in data processing, manually checking its results. I also used ChatGPT 5.2 (as well as Scopus and Google Scholar) to help find academic papers to support points in this document. I uploaded a complete draft to Gemini 3 and followed some of its minor suggestions for improvement.

## References

Bazerman, C. (1988). *Shaping written knowledge: The genre and activity of the experimental article in science*. Madison: University of Wisconsin Press.

Cheng, Q., Dai, Y., Liu, X., & Peng, S. (2026). The Trust Crisis in Artificial Intelligence: AI Hallucinations and Human-AI Collaboration. *Technology in Society*, 103286.

Demir, C. (2018). Hedging and academic writing: an analysis of lexical hedges. *Journal of language and linguistic studies*, *14*(4), 74-92.

Elsevier (2024). Insights 2024: Attitudes toward AI. https://www.elsevier.com/insights/attitudes-toward-ai

Hamidi, H., Sarem, S. N., & Lotfi, A. H. (2025). Exploring Hedging Devices in Scientific Research Papers: A Content Analysis Study of the 'Medical Journal of the Islamic Republic of Iran'. *Medical Journal of the Islamic Republic of Iran*, *39*, 125.

Hemminger, B. M., Lu, D., Vaughan, K. T. L., & Adams, S. J. (2007). Information seeking behavior of academic scientists. *Journal of the American society for information science and technology*, *58*(14), 2205-2225.

Herrera, G. (2011). Google Scholar Users and User Behaviors: An Exploratory Study. *College & Research Libraries*, *72*(4), 316-330.

Hyland, K. (1998). Hedging in scientific research articles. Amsterdam, The Netherlands: John Benjamins B.V.

Hyland, K. (2005). *Metadiscourse: exploring interaction in writing*. New York: Continuum.

Kuik, C. C. (2023). Shades of grey: Riskification and hedging in the Indo-Pacific. *The Pacific Review*, *36*(6), 1181-1214.

Kwon, D. (2024). Publishers are selling papers to train AIs-and making millions of dollars. *Nature News*, *636*(8043), 529-530.

Lenardič, J., & Fišer, D. (2021). Hedging modal adverbs in Slovenian academic discourse. *Slovenščina 2.0: empirične, aplikativne in interdisciplinarne raziskave*, *9*(1), 145-180.

Livytska, I. (2019). The use of hedging in research articles on applied linguistics. *Journal of language and cultural education*, *7*(1), 35-53.

Loi, C. K., & Lim, J. M. H. (2019). Hedging in the Discussion Sections of English and Malay Educational Research Articles. *GEMA Online Journal of Language Studies*, *19*(1).
López-Cózar, E. D., Orduña-Malea, E., Martín-Martín, A., & Ayllón, J. M. (2017). Google Scholar: the big data bibliographic tool. In *Research analytics* (pp. 59-80). Auerbach Publications.
Mirzapour, F. (2016). Gender differences in the use of hedges and first person pronouns in research articles of applied linguistics and chemistry. *International Journal of Applied Linguistics and English Literature*, *5*(6), 166-173.
Mur-Dueñas, P. (2021). There may be differences: Analysing the use of hedges in English and Spanish research articles. *Lingua*, *260*, 103131.
Nguyen Thi Thuy, T. (2018). A corpus-based study on cross-cultural divergence in the use of hedges in academic research articles written by Vietnamese and native English-speaking authors. *Social Sciences*, *7*(4), 70.
Radovanović, A., & Vojnović, D. V. (2023). Hedges In tourism and hospitality-related research articles. *BAS-British & American Studies*, *29*.
Rezanejad, A., Lari, Z., & Mosalli, Z. (2015). A Cross-cultural Analysis of the Use of Hedging Devices in Scientific Research Articles. Journal of Language Teaching and Research, 6(6), 1384-1392. https://doi.org/10.17507/jltr.0606.29
Strevens, M. (2020). The Knowledge Machine: How Irrationality Created Modern Science. New York: Liveright.
Thelwall, M. & Mohammadi, E. (2026). Can small and reasoning Large Language Models score journal articles for research quality and do averaging and few-shot help? Scientometrics. https://doi.org/10.1007/s11192-026-05585-2
Van Noorden, R., & Perkel, J. M. (2023). AI and science: what 1,600 researchers think. *Nature*, *621*(7980), 672-675.
Yao, M., Wei, Y., & Wang, H. (2023). Promoting research by reducing uncertainty in academic writing: a large-scale diachronic case study on hedging in Science research articles across 25 years. *Scientometrics*, *128*(8), 4541-4558.
Yu, Q., & Wen, R. (2022). A corpus-based quantitative study on the interpersonal functions of hedges in Chinese and German academic discourse. *Heliyon*, *8*(9).